\documentstyle[preprint,aps,epsf]{revtex}
\newcommand{\beq}{\begin{equation}}
\newcommand{\eeq}{\end{equation}}
\newcommand{\ds}{\displaystyle}
\begin{document}
\tightenlines
\draft

\title  {
 Can shadowing mimic the QCD phase transition?
         }
\author {
L.V.~Bravina,$^{1,2}$ E.E.~Zabrodin,$^{1,2}$ 
Amand Faessler,$^{1}$ C.~Fuchs$^{1}$ \\
  }
\address{$^1$ 
 Institute for Theoretical Physics, University of T\"ubingen,\\
 Auf der Morgenstelle 14, D-72076 T\"ubingen, Germany
         }
\address{$^2$ 
 Institute for Nuclear Physics, Moscow State University,
 119899 Moscow, Russia
         }

\maketitle

\begin{abstract}
The directed flow of protons is studied in the quark-gluon string
model as a function of the impact parameter for S+S and Pb+Pb 
reactions at 160 AGeV/$c$. A significant reduction of the directed 
flow in midrapidity range, which can lead to the development of the
antiflow, is found due to the absorption of early emitted particles
by massive spectators (shadowing effect). This effect can mimic
the formation of the quark-gluon plasma (QGP). However, in the
absorption scenario the antiflow is stronger for the system of
light colliding nuclei than for the heavy ones, while in the case
of the plasma creation the effect should be opposite.
\end{abstract}


\widetext

The search for a phase transition from nuclear matter to the
quark-gluon plasma (QGP) is one of the main goals of 
heavy ion collisions at ultrarelativistic energies. The creation of
the QGP at temperatures well above the transition temperature
$150\ {\rm MeV}\  \leq T \leq 200$ MeV is expected at energies of
heavy ion colliders, RHIC ($\sqrt{s} = 200$ AGeV) and LHC 
($\sqrt{s} = 5.5$ ATeV).
However, the possibility to reach the transition region of the mixed
hadron-QGP phase at AGS ($E_{lab} = 10.7$ AGeV) and SPS ($E_{lab} =
160$ AGeV) energies is not ruled out.

The collective flow of particles produced in heavy ion collisions 
is one of the few signals which are extremely sensitive to the 
formation of a new state of matter (for review see, e.g., i
\cite{StGr86,ReRi97,Oll98} and references therein),
since the flow is connected to the equation of state (EOS) of the
medium. The presence of the QGP undergoing a phase transition to 
hadronic matter results in a reduction of pressure \cite{Am91,Br94} 
and, therefore, causes the
so-called softening of the EOS \cite{HuSh95,Ri96} which can be
detected experimentally. This inspires a great 
interest of both experimentalists and theoreticians for the 
phenomenon of the transverse (to the beam axis) collective flow. The 
transverse flow is conventionally subdivided into radial, directed 
and elliptic flow. The latter two are attributed only to noncentral
collisions. Directed flow develops along the $x$-axis (impact
parameter axis perpendicular to the beam axis) of the reaction plane. 
The subject of the present
study is the directed flow of protons, which is most distinct
and less affected by other effects \cite{Vol97}. The directed flow of
protons is believed to have a characteristic $S$-shape of the mean 
transverse momentum $\langle p_x \rangle$ versus rapidity $y$ in the 
reaction plane. Indeed, this behaviour was observed in experiments at 
BEVALAC/SIS energies ($E_{lab} = 100$ AMeV - 1 AGeV) 
\cite{Gu84,Re84,Ra95,Le93}, and at AGS and SPS 
energies \cite{ags,na49,wa98}.
However, the midrapidity region $|y_{c.m.}| \leq 0.4$ is usually
not covered by the experimental apparatus. 

Investigations of the directed flow at AGS energies with the 
quark-gluon string model (QGSM), made by one of the authors some 
time ago \cite{Br95}, revealed the appearance of the antiflow, 
caused by the shadowing effect, mainly in the midrapidity region. 
The antiflow is found to be most pronounced in semicentral and 
peripheral collisions of light nuclei. For heavy systems, like 
Au+Au, at 10.7 AGeV the effect seems to be quite small.

It is worth noting that the antiflow signal is not a feature 
attributed solely to the particular microscopic model such as QGSM.
Strong antiflow of protons in the midrapidity range is predicted by 
the VENUS model for very peripheral ($b = 8-10$ fm) Pb+Pb collisions
at SPS energy \cite{wa98}. The plateau in midrapidity region of the
directed flow in semiperipheral ($b \leq 7$ fm) Au+Au collisions at
$E_{lab} = 25$ AGeV is found in the UrQMD calculations \cite{Brac99}.
Studies of the nucleon directed flow in
terms of the first Fourier coefficient \cite{PoVo98}, $v_1$, 
of the particle azimuthal distribution, made recently within the
RQMD model for semiperipheral and peripheral heavy ion collisions at 
energies from SPS to RHIC \cite{LPX99,Sn99}, also demonstrate the 
anisotropy indicated by deviations of $v_1$ from the straight line 
behaviour in the midrapidity region.  

On the other hand, hydrodynamic simulations of heavy ion collisions
with the QGP EOS indicate the creation of a similar antiflow in the 
midrapidity region \cite{Br94,Brac99,Brac97,CsRo99}. No deviations 
from the straight line of the function $\langle p_x (y) \rangle $
in this range have been found in the simulations with a pure hadronic 
EOS. To get a clear signal of the plasma creation it is very 
important, therefore, to trace to what extent the development of the 
antiflow may be caused by any other processes not directly related to 
the formation of the QGP. We aim to study the dependence of the
directed flow of protons on the impact parameter in the collisions
of light and heavy ions at SPS energies.

For the simulations of nuclear interactions the microscopic 
quark-gluon string model with rescatterings \cite{qgsm} is employed. 
It treats the hadronic and nuclear interactions on the basis of 
the Gribov-Regge theory accomplished by the string phenomenology. 
As independent degrees of freedom the model includes octet and
decuplet baryons, and octet and nonet vector and pseudoscalar 
mesons, as well as their antiparticles. For simplicity the fine 
tuning mechanisms such as mean fields, enhanced in-medium cross 
sections, colour coherent effects or string-string interactions are 
disregarded in the present version of the model. A formation of the 
QGP is also not assumed in the QGSM, although one can consider the 
appearance of non-particle objects, strings, as a precursor of the 
plasma creation.
The model has successfully predicted the magnitude of the directed 
flow at SPS energies \cite{Am91} far before its experimental 
discovery \cite{na49,wa98}, and up to now it provides one of the 
best agreements with the experimental data \cite{wa98} among the 
microscopic models.

For the simulations at 160 AGeV two symmetric nuclear systems, a 
light ($^{32}$S+$^{32}$S) and a heavy one ($^{208}$Pb + $^{208}$Pb), 
have been chosen. The mean in-reaction plane transverse flow of 
protons as a function of rapidity is defined in the calculations as
\beq
\ds
\langle p_x(y)/A \rangle = p_x \frac{d N^p}{d y} \left/ 
\frac{d N^p}{d y} \equiv \left \langle \left(
\vec{p}_T(y)/N^p,\vec{b} 
\right) / |\vec{b}| \right \rangle \right. ,
\label{eq1}
\eeq
where $N^p$ is the number of protons and $b$ is the impact parameter.

To compare heavy ion systems with different radii, the distributions
(\ref{eq1}) are studied as a function of the reduced impact parameter, 
$\tilde{b} = b/b_{max}$. This parameter varies in the simulations 
from $\tilde{b} = 0.15$ (or even 0.05) for central collisions up to
$\tilde{b} = 0.9$ corresponding to rather peripheral ones. For a
symmetric system the 
maximum impact parameter is, apparently, $b_{max} = 2\, R_A$, that
gives $b_{max} = 7.11\, (13.27)$ fm for S+S (Pb+Pb) collisions.

\begin{figure}[htp]
\centerline{\epsfysize=18cm \epsfbox{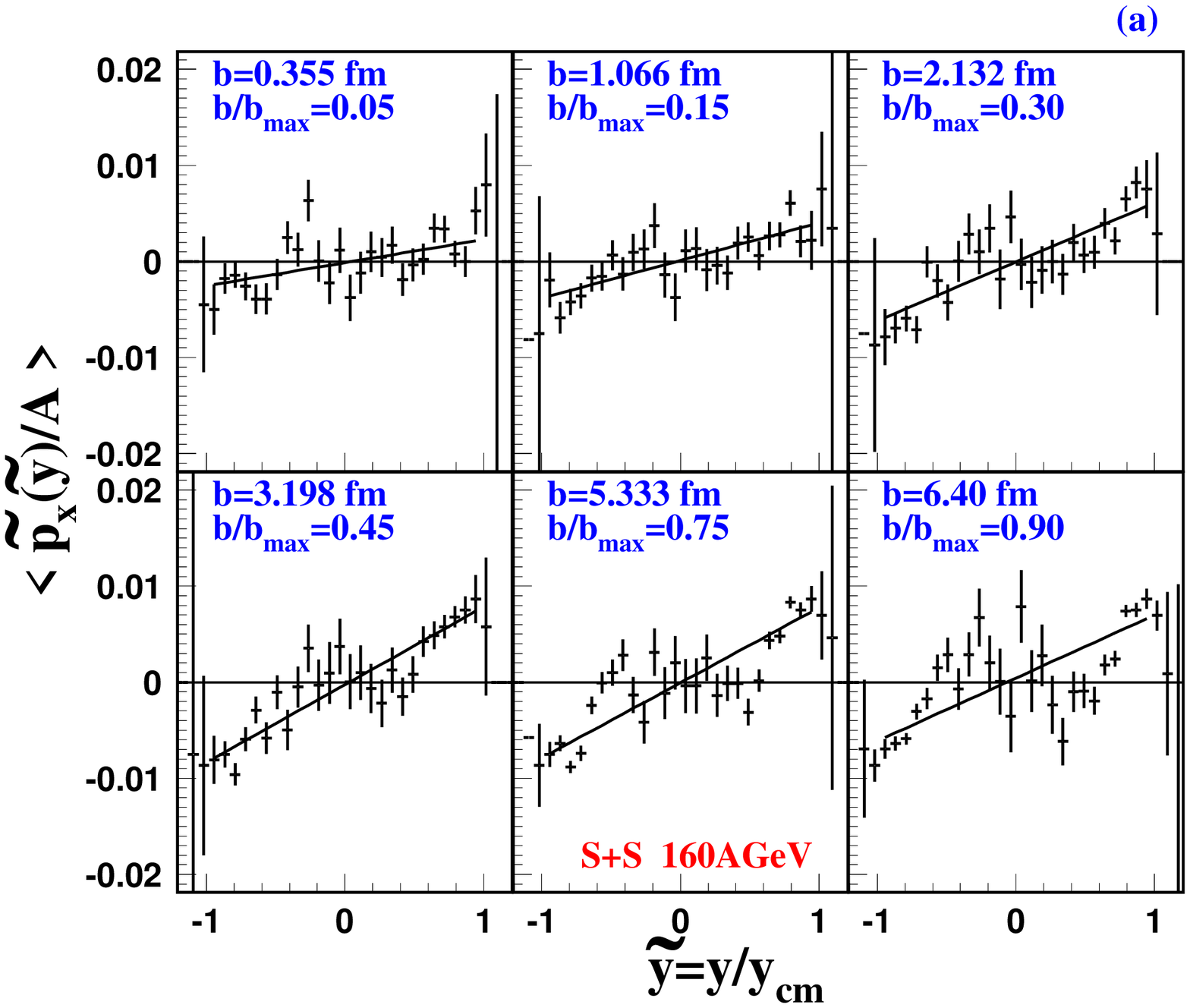}}
\caption{
Directed flow of protons, $\langle \tilde{p}_x(\tilde{y})/A 
\rangle $, as a function
of rapidity, $y$, calculated in QGSM for $^{32}$S+$^{32}$S (a) and
$^{208}$Pb+$^{208}$Pb (b) calculations at 160 AGeV. Values of the 
impact parameter, $b$, and the reduced impact parameter, $\tilde{b} 
= b/b_{max}$, are listed in each panel of the figure. 
Results of the simulations are fitted to linear dependence (solid
line) in rapidity interval $|\tilde{y}| \leq 1$.
}
\centerline{\epsfysize=18cm \epsfbox{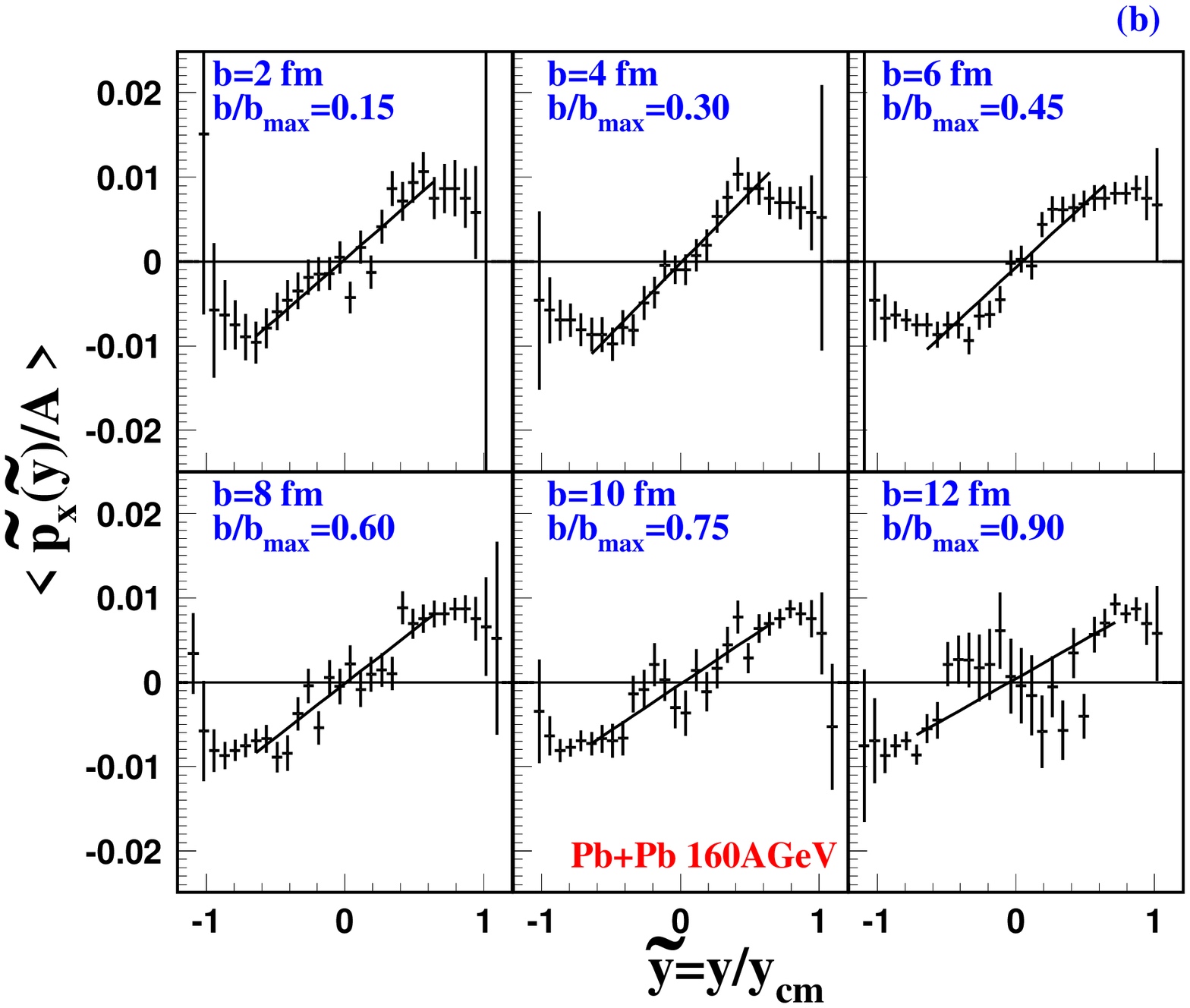}}
\label{fig1}
\end{figure}

Results are shown in Figs.~\ref{fig1}(a) and \ref{fig1}(b). 
Here the in-plane transverse momentum of protons is defined in the
scale-invariant form \cite{Cs94} $\langle \tilde{p}_x (\tilde{y})/A
\rangle$, where the reduced momentum, $\tilde{p}_x = p_x/p_{c.m.}$,
and reduced rapidity, $\tilde{y} = y/y_{c.m.}$, are normalized to
the center-of-mass momentum and rapidity of the projectile,
$p_{c.m.} = 8.65$ and $y_{c.m.} = 2.9$, respectively.
Immediately one can see that for the light system 
[Fig.~\ref{fig1}(a)] the flow has a more complex structure in the 
midrapidity region compared to a simple linear dependence shown by
a solid line in Fig.~\ref{fig1}. The antiflow in the central 
rapidity region $|y| \leq 1$ develops already at $\tilde{b} \leq 0.3$. 
Towards the semicentral and peripheral collisions the zone of 
irregularities in the behaviour of the directed flow becomes broader. 
In the case of heavy nuclei [Fig.~\ref{fig1}(b)] the disappearance of 
the flow occurs also at $\tilde{b} = 0.3$ but within the narrower 
region, $|y| \leq 0.25$. The distinct antiflow signal seems to appear 
only in peripheral, $\tilde{b} \geq 0.75$, collisions. 
As expected, the flow signal at an SPS energy is significantly weaker
compared to that in calculations at an AGS energy \cite{Br95}.

To quantify the strength of the antiflow and normal flow separately
the reaction zone was subdivided into rectangular cells with
volumes $V^{cell} = 1.6 \times 1.6 \times 1.2 = 3$ fm$^3$. The
time evolution of the baryon density, $n_{\rm B}$, and the 
collective velocity of the cells, $\vec{v}_{cell}$, was considered. 
To distinguish between the normal flow and antiflow components the 
step function, $\Theta$, of the product $p_x^{cell} y^{cell}$ is
used in the calculations:\\

$\Theta_{flow} = \left\{ \begin{array}{ll}
  1, & \mbox{if $p_x^{cell} y^{cell} > 0$} \\
  0, & \mbox{otherwise ;}  \end{array}
\right.$\\

$\Theta_{a.-flow} = \left\{ \begin{array}{ll}
  1, & \mbox{if $p_x^{cell} y^{cell} < 0$} \\
  0, & \mbox{otherwise .}  \end{array}
\right.$\\

\begin{figure}[htp]
\centerline{\epsfysize=18cm \epsfbox{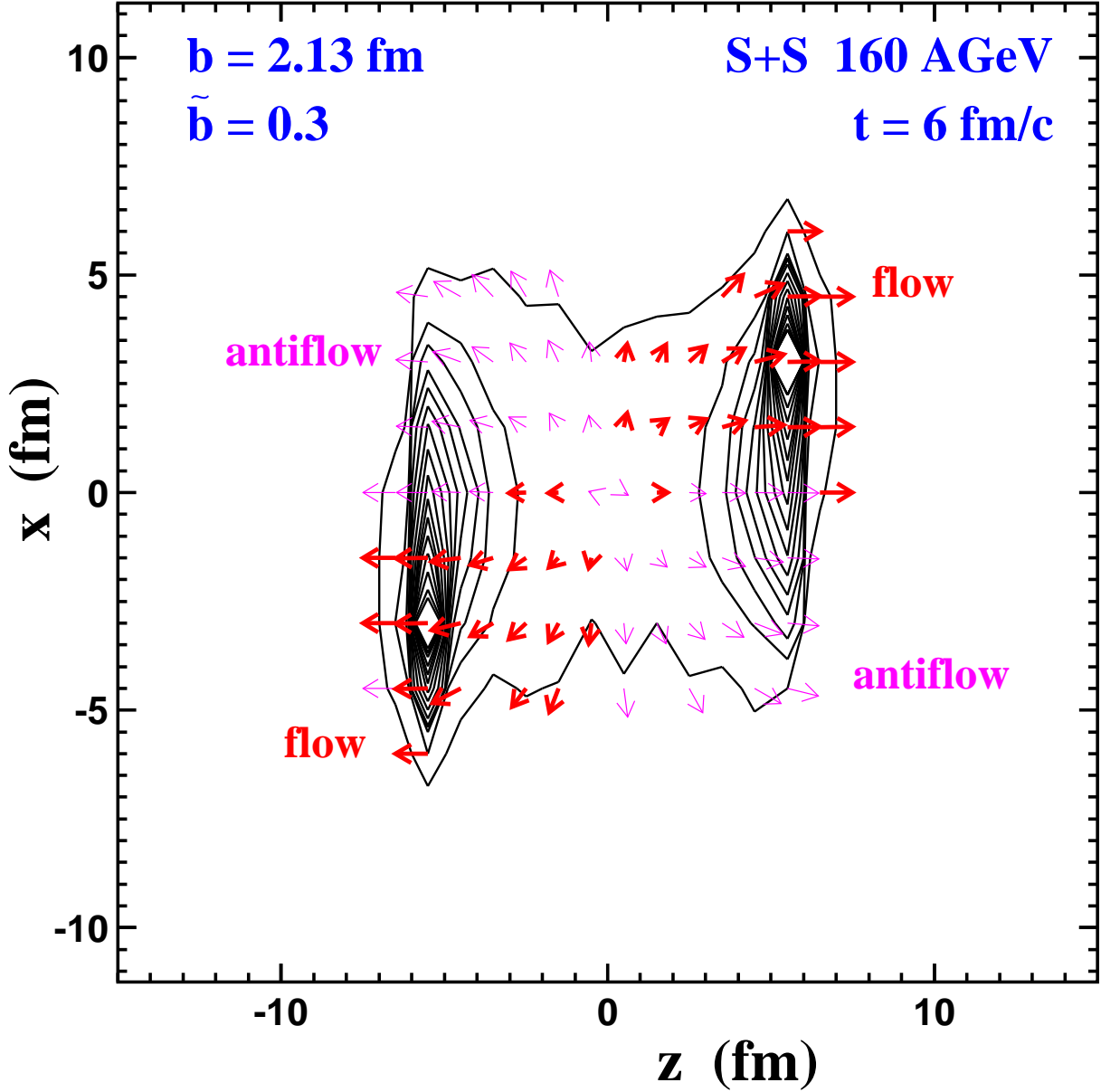}}
\caption{
Space-time evolution of the baryon density, $n_{\rm B} / n_0$ 
(contour plots), and collective velocity, $\vec{v}$ (arrows), of
the cells with volume $V = 3$ fm$^3$ each. Calculations are made
in QGSM for semicentral ($b = 2.13$ fm) S+S collisions at SPS
energy for all formed baryons at time $t = 6$ fm/$c$. Contour plots 
correspond to $n_{\rm B} / n_0 = 0.05,\ 0.25,\ 0.5,\ etc.$ of 
normal baryon density.
}
\label{fig2}
\end{figure}

Figure~\ref{fig2} depicts a snapshot of baryon densities and 
collective velocities of the cells in S+S collisions at SPS energy,
calculated at fixed impact parameter $b = 2.13$ fm $(\tilde{b} = 
0.3)$ at time $t = 6$ fm/$c$. The antiflow component develops towards 
the relatively more dilute zones of nuclear collisions, whereas the
normal flow component follows the baryon rich remnants of the 
interacting nuclei. Apparently, the total flow is
determined by the interplay between these two components. This leads
to the reduction of the total flow because of their mutual 
cancellation.

\begin{figure}[htp]
\centerline{\epsfysize=18cm \epsfbox{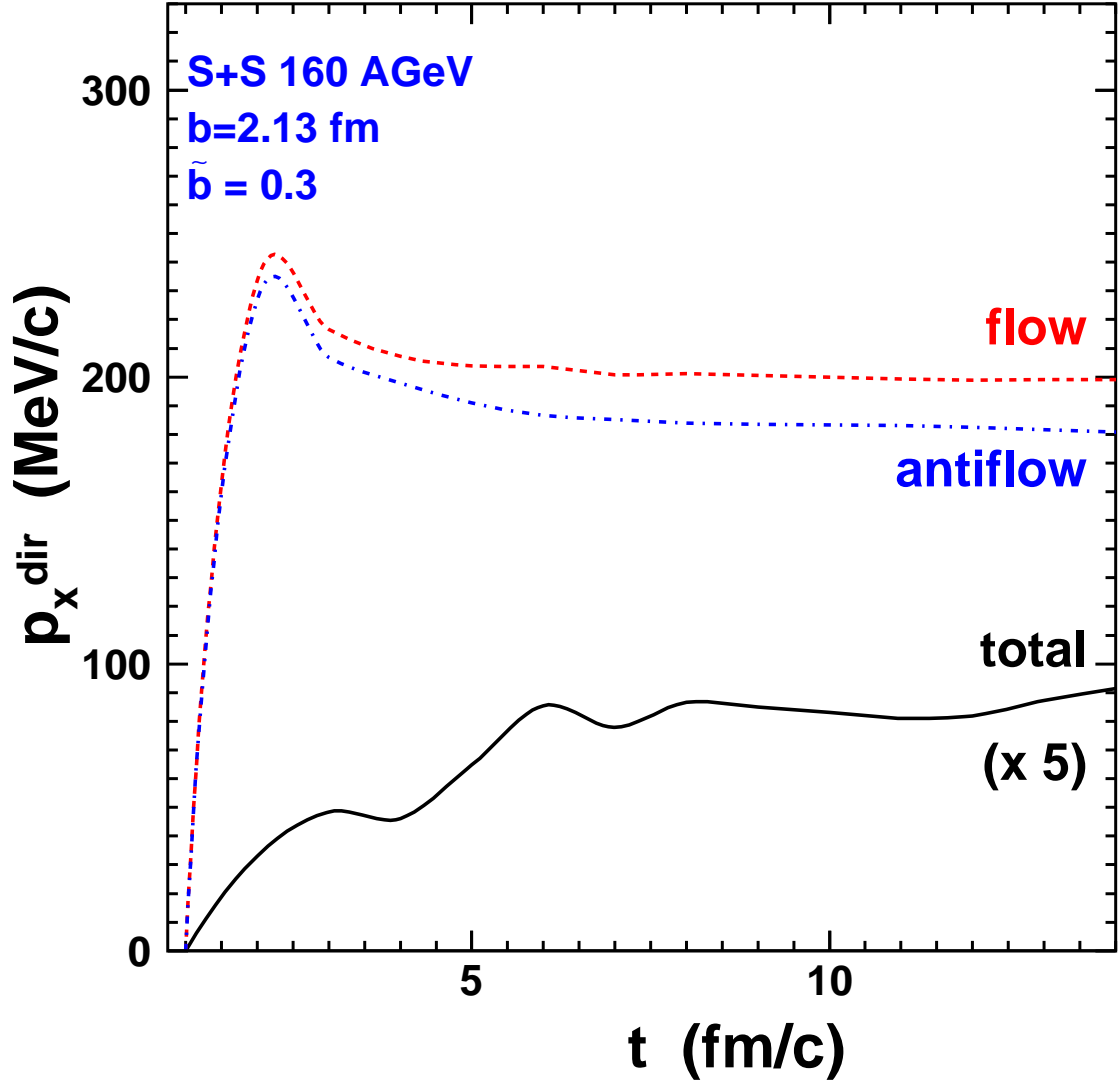}}
\caption{
Time evolution of the mean directed flow of protons in S+S 
collisions with the reduced impact parameter $\tilde{b} = 0.3$ at 
160 AGeV/c. Dashed and dash-dotted curves indicate the normal and 
antiflow components, respectively. Full curve denotes the resulting 
flow multiplied by factor 5.
}
\label{fig3}
\end{figure}

To calculate the mean directed flow of protons in the forward or
backward hemisphere of the center-of-mass system, 
$\langle p_x^{dir}/N^p \rangle$, the proton rapidity distribution,
$d N^p / d y$, should be integrated over the whole rapidity 
range with the weight $\langle p_x (y) / N^p \rangle$, given by 
Eq.~(\ref{eq1}), namely,
\beq
\ds
\left \langle \frac{p_x^{dir}(y)}{N^p} \right \rangle = \int 
{\rm sign}(y) 
\frac{d N^p}{d y} \left \langle \frac{p_x(y)}{A} \right \rangle dy 
\left/ \int \frac{d N^p}{d y} dy \right. .
\label{eq2}
\eeq
The time evolution of the proton directed flow in forward hemisphere 
is shown in Fig.~\ref{fig3}. We see that both, the normal flow 
component and the antiflow component, develop quickly and reach a
maximum at $t \cong 2$ fm/$c$ after the beginning of collision.
The partial flows are almost two orders of magnitude larger than the
resulting directed flow. It is worth noting also that the total
directed flow of protons decreases slightly after $t = 6$ fm/$c$.
This is due to decays of baryonic resonances, which dominate over 
the formation of resonances at the late stage of the reaction.
Despite the normal flow component, integrated over the whole rapidity
range, is always slightly larger than the antiflow one, in some
rapidity windows the antiflow component can overshadow its normal
counterpart. For instance, particles emitted with small rapidities at 
the early stage of the reaction in ``normal" direction are absorbed by 
two massive spectators, while particles emitted in the opposite 
direction remain unaffected. The signal becomes weaker with increasing 
centrality of the collision because of the shrinking of space where 
the particles could be emitted unscreened.
In heavy systems the emission of particles from the central fireball
increases. Thus the absorption of several early emitted particles by
the flying spectators becomes less effective, and the relative 
strength of the antiflow decreases.

\begin{figure}[htp]
\centerline{\epsfysize=18cm \epsfbox{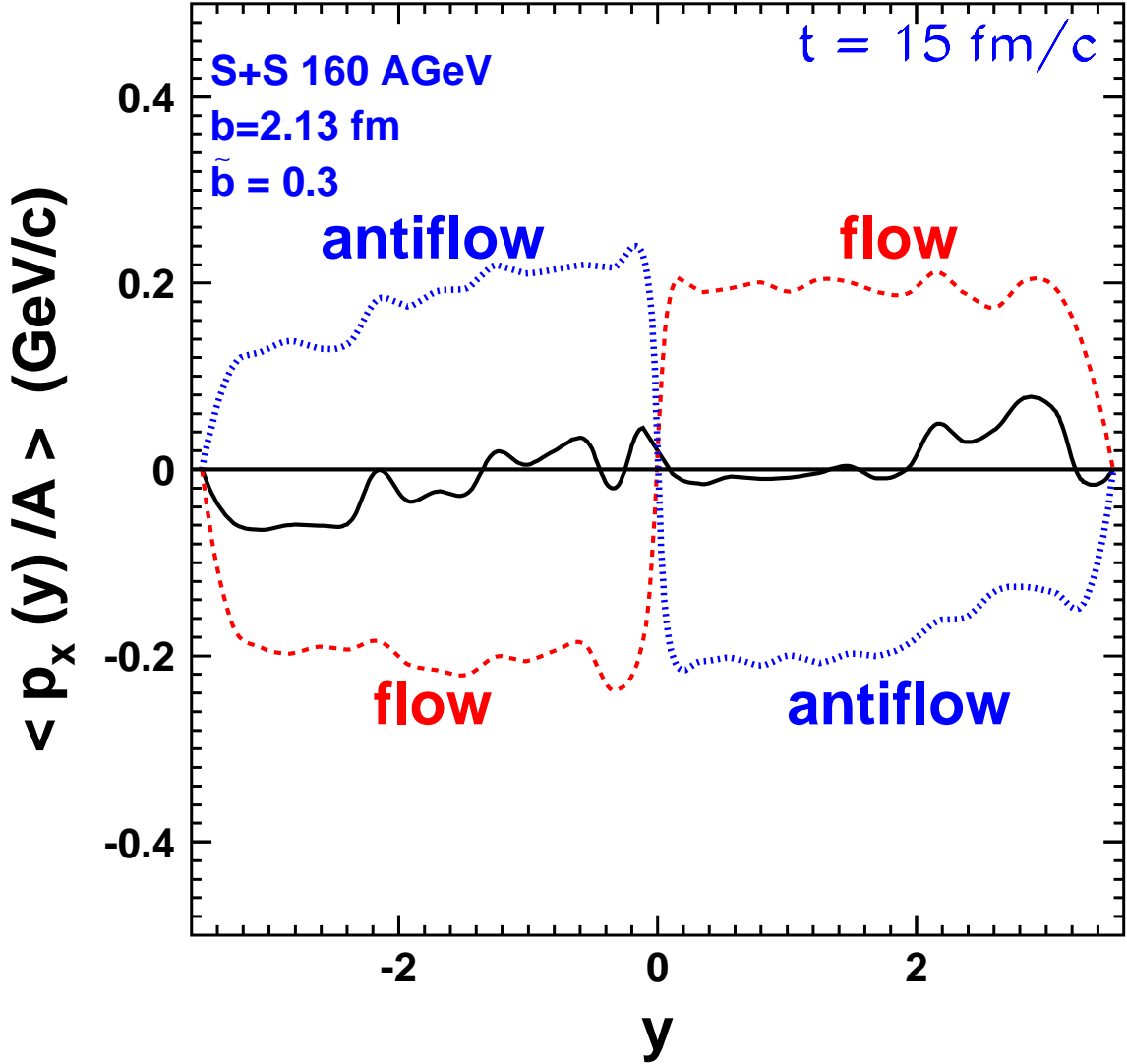}}
\caption{
Rapidity distribution of the directed flow of protons at $t=15$
fm/$c$ in S+S collisions with $\tilde{b} = 0.3$ at SPS energy.
Dashed, dotted, and full curves indicate the normal and antiflow 
components, and the resulting flow, respectively. 
}
\label{fig4}
\end{figure}

Figure~\ref{fig4} presents the rapidity 
distribution of the directed flow of protons together with the flow
and antiflow components at the late stage of S+S collisions with 
$\tilde{b}=0.3$. The resulting flow has a characteristic wavy 
structure. It tends generally to
grow with rapidity rising from $y = -2.5$ to $y = -0.4$, and from
$y = 0.4$ to $y=2.5$. In the midrapidity range $|y| \leq 0.4$ the
total flow decreases. This effect can imitate the expected softness 
of the EOS due to the plasma creation and, therefore, it should be 
taken into account in the analysis of experimental data.

In conclusion, directed flow of protons is studied in semicentral 
and peripheral sulphur-sulphur and lead-lead collisions at 160 AGeV
generated by a quark-gluon string model of nuclear collisions. The 
flow is shown to have a complex structure in the rapidity range 
$|y| \leq 1.5$, which is strongly dependent on the centrality of
the collision, as well as on the mass number of colliding nuclei.
The total directed flow can be decomposed onto the normal component,
which follows the outgoing residues of collided ions, and the 
antiflow component, which develops in the opposite direction, where
the baryon matter is more dilute.

Although these partial components are relatively large, their
mutual cancellation leads to a rather modest signal of the total
flow. In S+S collisions with an impact parameter $b \cong 2$ fm
$(\tilde{b} \cong 0.3)$ the antiflow already dominates over the 
normal flow in the central rapidity range $|y| \leq 0.4$. This
circumstance causes the decrease of the total flow. In heavy 
systems like Pb+Pb the effect is most pronounced in peripheral
collisions with $b \cong 9$ fm $(\tilde{b} \cong 0.7)$ or larger,
although the significant disappearance of the flow in the range
$|y| \geq 0.5$ takes place already in semicentral events with 
$b = 4$ fm ($\tilde{b} = 0.3$). This can be misinterpreted as an
indication for the softening of the EOS of strongly interacting
matter. 

Note that in the microscopic string model like QGSM the total 
directed flow of protons changes its behaviour in the central
rapidity region because of almost purely geometrical reasons.
Therefore, the effect is more distinct in the collisions of light
ions. With the rise of incident energy the remnants of nuclei move 
faster thus giving space for the antiflow development. Hence, the 
characteristic change of sign of the total flow in heavy ion 
collisions should appear not only in peripheral collisions, but 
also in semicentral and even in almost central collisions (1 fm 
$ \leq b \leq $ 3 fm) at RHIC energies $(\sqrt{s} = 200$ AGeV).

This description is opposite to the picture of non-central heavy
ion collisions given by hydrodynamic models. In hydrodynamics the
creation of a QGP is likely to occur in heavy (Pb+Pb or Au+Au) 
systems rather than in light (S+S) systems. The formation even of 
a small amount of plasma with the subsequent phase transition to 
the hadron phase can also produce the negative slope of the 
$\langle p_x(y) \rangle$-distribution in the midrapidity region. 
The effect, therefore, should essentially be more pronounced in 
heavy ion collisions compared to collisions of light ions with the 
same reduced impact parameter. The situation clearly awaits better
experimental data on proton directed flow in the central rapidity
region.   

{\bf Acknowledgments.} 
We acknowledge fruitful discussions with L. Csernai, M. Gyulassy, 
E. Shuryak, H. Sorge, H. St\"ocker, and Nu Xu.
This work was supported in part by the Bundesministerium f\"ur 
Bildung und Forschung (BMBF) under contract 06T\"U887.

\widetext

\newpage

\end{document}